\documentclass[manuscript]{aastex631}
\renewcommand{\sun}[0]{\odot}

\newcommand{\avg}[1]{\left< #1 \right>} 
\begin{document}

\title{Evidence for Parameteric Decay Instability in the Lower Solar Atmosphere}

\correspondingauthor{Michael Hahn}
\email{mhahn@astro.columbia.edu}

\author{Michael Hahn}
\affiliation{Columbia Astrophysics Laboratory, Columbia University, 550 West 120th Street, New York, NY 10027}

\author{Xiangrong Fu}
\affiliation{New Mexico Consortium, 4200 W. Jemez Rd, Suite 200, Los Alamos, NM 87544}

\author{Daniel Wolf Savin}
\affiliation{Columbia Astrophysics Laboratory, Columbia University, 550 West 120th Street, New York, NY 10027}

\begin{abstract}

We find evidence for the first observation of the parametric decay instability (PDI) in the lower solar atmosphere. Specifically, we find that the power spectrum of density fluctuations near the solar transition region resembles the power spectrum of the velocity fluctuations, but with the frequency axis scaled up by about a factor of two. These results are from an analysis of the Si~\textsc{iv} lines observed by the \textit{Interface Region Imaging Spectrometer (IRIS)} in the transition region of a polar coronal hole. We also find that the density fluctuations have radial velocity of about 75~$\mathrm{km\,s^{-1}}$ and that the velocity fluctuations are much faster with an estimated speed of $250$~$\mathrm{km\,s^{-1}}$, as is expected for sound waves and Alfv\'en waves, respectively, in the transition region. Theoretical calculations show that this frequency relationship is consistent with those expected from PDI for the plasma conditions of the observed region. These measurements suggest an interaction between sound waves and Alfv\'en waves in the transition region that is evidence for the parametric decay instability.

\end{abstract}

\section{Introduction} \label{sec:intro}

Coronal holes are open magnetic field regions that are known to be the source of the fast solar wind. One of the heating mechanisms of these regions is theorized to occur through Alfv\'en wave turbulence \citep[e.g.,][]{Suzuki2006, Hollweg2007, Cranmer2007}. The basic picture is that Alfv\'en waves are excited at the base of the corona and travel outward along the open field lines. Some waves are reflected off gradients in the Alfv\'en speed. A nonlinear interaction between the outward and reflected Alfv\'en waves leads to Alfv\'enic turbulence \citep{Howes:PoP:2013}, which drives energy to small length scales where the energy can go into plasma heating. This picture has been supported by observations showing that Alfv\'en waves do dissipate at low heights in coronal holes \citep{Bemporad:ApJ:2012, Hahn2012, Hahn:ApJ:2013, Hara:ApJ:2019}. However, theoretical models suggest that the background magnetic field and density gradients in coronal holes are too weak to generate sufficient reflection and turbulence \citep{Asgari:ApJ:2021}. The heating rate can be increased by the addition of small scale gradients due to density fluctuations \citep{vanB2016, vanB2017}. \citet{Asgari:ApJ:2021} recently incorporated observationally constrained density fluctuations \citep{Miyamoto:ApJ:2014, Hahn:ApJ:2018} into a wave-turbulence heating model and showed that the observed level of density fluctuations causes enough Alfv\'en wave reflection and turbulence to heat coronal holes. It is surprising to find such large amplitude density fluctuations in the corona, as they are expected to be efficiently damped \citep{Ofman:ApJ:1999, Ofman:ApJ:2000}. This raises the question as to the origin of the observed density fluctuations. 

One possibility is that the density fluctuations are produced through a nonlinear interaction with the Alfv\'en waves, known as the parametric decay instability \citep[PDI;][]{golds_apj_1978,derby_apj_1978}. Theoretical models and computer simulations have shown that in a low-$\beta$ plasma where the magnetic pressure dominates the fluid pressure, such as the solar corona, a large amplitude forward propagating Alfven wave can decay into a backward propagating Alfven wave and a forward propagating ion acoustic wave, self-consistently generating density fluctuations leading to turbulence and heating \citep[e.g.,][]{Chandran:JPP:2018, fu_apj_2018, Reville:ApJ:2018, Shoda:ApJ:2019}. We are unaware of any observations to date that have  seen direct evidence for PDI at the Sun. 


Here, we investigate the relationship between density fluctuations and velocity fluctuations transverse to the mean magnetic field in the solar transition region. We use observations from the \textit{Interface Region Imaging Spectrograph} \citep[\textit{IRIS};][]{DePontieu:SolPhys:2014} and analyze data from the Si~\textsc{iv} lines. Intensity fluctuations represent changes in the density, which are consistent with acoustic waves.  Doppler shifts of the spectral lines indicate velocity fluctuations, which are likely due to Alfv\'enic waves. We have found that the power spectrum of the density and velocity fluctuations are similar to one another except for a scaling factor of the frequency axis. This property suggests that we are observing an interaction between Alfv\'enic and acoustic waves through PDI \citep{Sagdeev:1969, derby_apj_1978, golds_apj_1978}. Other measured plasma parameters and theoretical calculations of the instability growth rate are also consistent with PDI. 

The rest of this paper is organized as follows. In Section~\ref{sec:obs} we describe the observations. Details of the analysis are presented in Section~\ref{sec:anal}. We then discuss the implications of the results with comparisons to theoretical predictions for PDI in Section~\ref{sec:discuss}. Some alternative hypotheses are also presented there. Section~\ref{sec:conclusions} concludes. 

\section{Instrument and Observations}\label{sec:obs}

\begin{figure}[h]
	\centering \includegraphics[width=0.8\textwidth]{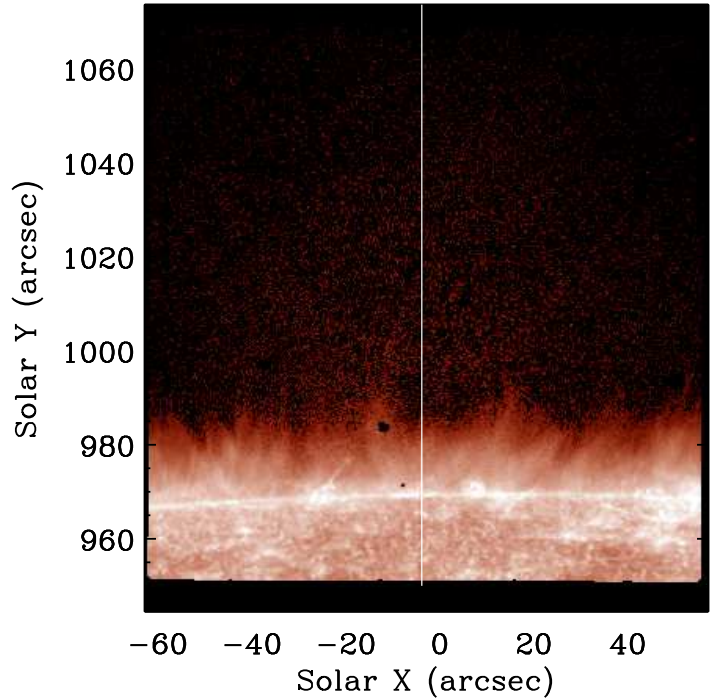}
	\caption{\label{fig:iris_sj} \textit{IRIS} slit-jaw image for the observed region. The position of the slit is highlighted by the vertical line. 
	}
\end{figure}

We studied an \textit{IRIS} observation of the off-limb transition region. This was a sit-and-stare type observation starting at 2016-10-31 19:45 UT, where the slit was positioned at $x=-4^{\prime\prime}$ relative to the central meridian of the Sun. The slit extended from $y=944.9$--$1073.6^{\prime\prime}$ in the vertical direction relative to Sun center. At the time, the solar radius $R_\sun$ was at $966.8^{\prime\prime}$. So, the observation covered the region where $0.977 R_{\sun}<r<1.110 R_{\sun}$. Data were taken at a cadence of $\Delta t = 9.34$~s for a total time interval of $\approx 4600$~s. Figure~\ref{fig:iris_sj} shows the slit-jaw image obtained by \textit{IRIS} in the 1400~\AA\ bandpass, which is dominated by emission from Si~\textsc{iv}. The position of the slit for the spectroscopic data is indicated in the figure. 

\textit{IRIS} level 2 spectral data were processed using the standard \textit{iris\_prep} routine \citep{Wulser:SolPhys:2018, IRISUG}. Orbital variations in the wavelength axis were corrected using the methods described in \citet{IRIS:wavelengthcal} and \citet{Wulser:SolPhys:2018}, which assume the photospheric lines are unshifted. At this point the wavelength scale has been converted to physical units, but the intensity scale remains in data number (DN) units. We worked with these data and fit the spectral lines with a single Gaussian in order to extract the intensity, wavelength, and line width for all lines of interest. Inspection of the fits demonstrated that single Gaussians fit the data well and there did not appear any advantage to using double Gaussians. There are, however, some suggestions that multiple emission components might be present in the analysis at a level that is difficult to resolve, as we discuss below. For analysis of plasma properties where an absolute intensity calibration is needed, we applied the radiometric calibration described in \citet{IRIS:StellarCalibration} and \citet{Wulser:SolPhys:2018} to convert the fitted intensity in DN to physical units. 

Our analysis focused on the Si~\textsc{iv} lines at 1394 and 1403~\AA. All of the analysis was repeated for both lines with consistent results; but as the 1394~\AA\ line is brighter, we present results mainly for that line. The ratio of intensities of these lines gives a measure of the optical thickness of the observation, with an optically thin plasma having a line intensity ratio of $I_{1394}/I_{1403} \approx 2$.  We found that the data just above the limb were somewhat optically thick with the ratio approaching the optically thin limit at about $y=980^{\prime\prime}$. However, in the analysis we do not find any significant systematic effects due to the varying optical thickness. 

For the analysis of waves and fluctuations, we focused on 64 pixels in the off-limb region spanning from $y=972^{\prime \prime}$ to $y=993^{\prime\prime}$. This height range was chosen to avoid some of the complex structures and spicules at the lowest heights. Larger heights were not useful for the analysis as the Si~\textsc{iv} line intensity drops rapidly and the data there are dominated by noise. 

\section{Analysis and Results}\label{sec:anal}

\subsection{Fluctuation Power Spectra}\label{subsec:power}

\subsubsection{Density Fluctuations}\label{subsubsec:dnpower}

Density fluctuations $\delta n$ are observed as intensity fluctuations, $\delta I$. For a collisionally excited plasma, such as analyzed here, $I \propto n_{\mathrm{e}}^2$ and so $\delta n_{\mathrm{e}}/n_{\mathrm{e}} = \delta I / 2 I$. 
In order to analyze the power spectrum of the intensity fluctuations we first subtract the average background intensity level and any long-period trend by fitting the $I(t)$ data for each spatial pixel along the \textit{IRIS} slit with a linear function. From this we obtained a stationary time series. We then found the deviations from the linear fit, $\delta I(t,y)$. To control for the radial variation of the average intensity with height, we normalize these intensity fluctuations by the time-averaged intensity at each height $\avg{I}(y)$ to obtain $\delta I (t,y) / \avg{I}(y)$. For brevity, we will refer to this quantity as the intensity fluctuation with the symbol $\delta I(t,y)$. In magnitude, the typical root-mean-square (RMS) intensity fluctuation level was about 16\%, so the density fluctuation amplitudes are $\delta n_{\mathrm{e}}/n_{\mathrm{e}} \approx 8\%$. 

\begin{figure}[t]
	\centering \includegraphics[width=0.8\textwidth]{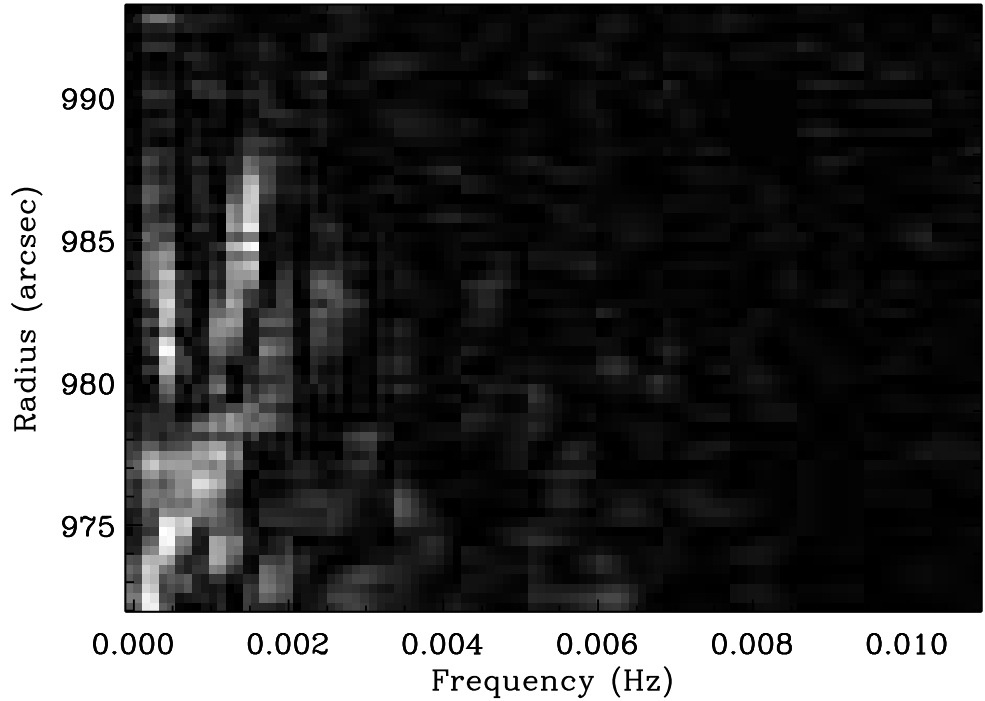}
	\caption{\label{fig:Ipsd_vs_y} Power spectrum of the relative density fluctuations as a function of height. Here the power spectrum is normalized by the total power at each height. The color scale is linear with a maximum value of 0.15 shown in white and zero in black. 
	}
\end{figure}

Figure~\ref{fig:Ipsd_vs_y} illustrates the Fourier power spectrum for the intensity fluctuations as a function of height. A bootstrap method was used to quantify the significance of the power spectrum peaks \citep{Nemec:AJ:1985}. For each pixel in the data we have a set of 494 intensities $\delta I_{j}$, each corresponding to time $t_j$. The periodicity reflected by the power spectrum occurs only when the set of all the observed intensities $\delta I_j$ occurs in the observed order. In order to test the significance of the peaks, we used the same set of intensities but scrambled the ordering. The result is a data series that represents random noise with the same properties of the observed $\delta I_j$, but no periodicity. We then performed the power spectrum analysis on each noise permutation. This was repeated for 200 random permutations. As a result, we obtain a histogram of the noise power distribution at each frequency. The significance of the power is determined by comparing the measured power at each frequency to the noise power histogram at that same frequency. For the intensity fluctuations, the peaks in the power spectrum below 10~mHz are significant at the 95\% level or better. That is, there is a less than 5\% chance that those peaks arise due to random noise. 

The magnitude of each peak has, in principle, an uncertainty of 100\% \citep{Press:NR}. This is because for a time series with $N$ data points, the Fourier analysis determines the amplitude and phase at $N/2$ frequencies. In order to reduce the uncertainty it is necessary to average the power spectra. For example, a common choice is to bin  the power spectrum in frequency. In order to preserve frequency resolution, we chose instead to average the power spectrum over the observed height range. This relies on the assumption that the power spectrum is not varying systematically with height. This appears to be the case based on Figure~\ref{fig:Ipsd_vs_y}, where although there is some structure in the power spectrum versus height there is no clear trend. As our analysis was performed over 64 vertical spatial pixels, there are 64 measurements of the power spectrum, so the uncertainty in the average power spectrum is reduced to about 13\%. 

\subsubsection{Velocity Fluctuations}\label{subsubsec:velocity}

Velocity fluctuations $\delta v$ are observed as Doppler shifts of wavelength, $\delta \lambda$. We obtained a stationary time series by fitting the $\lambda(t)$ at each height, $y$, with a linear trend and determined the fluctuation relative to that trend. For the power spectrum analysis of velocity fluctuations, we work directly with the wavelength data, $\delta \lambda(t,y)$, as the absolute magnitude of the power spectral peaks is not important. The typical RMS amplitude of the wavelength fluctuations was about 0.012~\AA. Since $\delta v = c \delta \lambda/\lambda_{0}$, where $c$ is the speed of light and $\lambda_{0} = 1394$~\AA, this corresponds to an RMS velocity amplitude of about 2.6~$\mathrm{km\,s^{-1}}$. However, previous studies have shown that Doppler shift measurements of velocities underestimate the actual wave amplitudes due to the line-of-sight integration \citep{McIntosh:ApJ:2012} and so are considered to provide a lower limit for the wave amplitudes. The Doppler shift measurements reveal the relative power spectrum, but absolute wave amplitudes are better estimated using line widths (see Section~\ref{subsec:width}). 

\begin{figure}[t]
	\centering \includegraphics[width=0.8\textwidth]{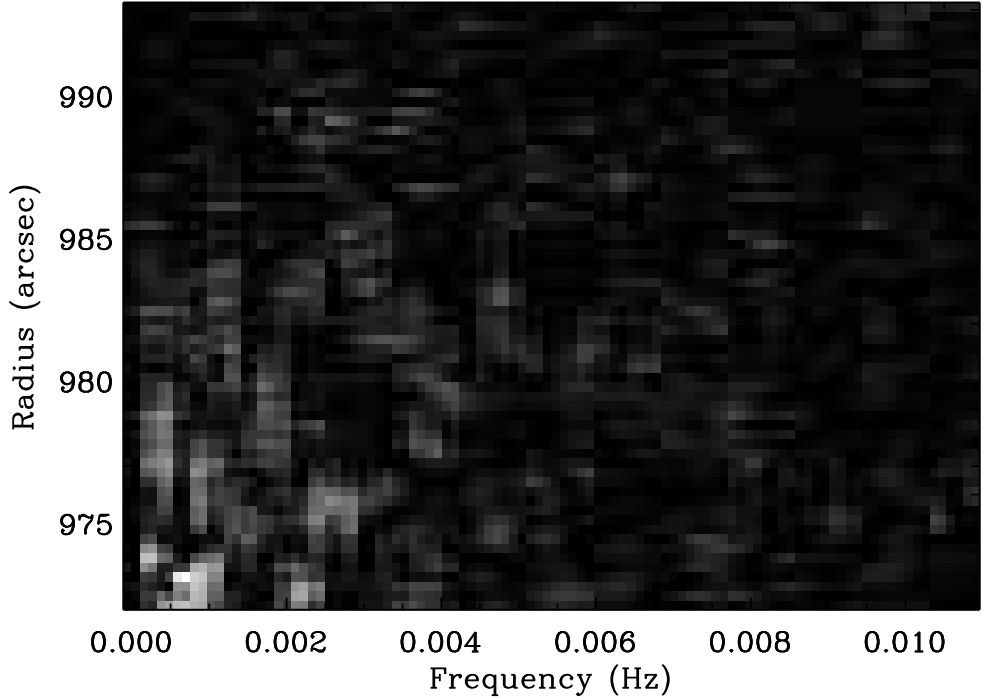}
	\caption{\label{fig:Vpsd_vs_y} Same as Figure~\ref{fig:Ipsd_vs_y}, but for the relative velocity fluctuations. The color scale is linear with a maximum value of 0.12. 
	}
\end{figure}

We obtained the Fourier power spectrum for $\delta \lambda(y,t)$ at each height using the same methods as for the intensity fluctuations (Figure~\ref{fig:Vpsd_vs_y}). Again, there did not appear to be a trend in the power spectrum with height. So, the average velocity fluctuation power spectrum was found by normalizing the power spectrum at each height by the total power at that height and then averaging the normalized spectra over the studied height range. 

\subsection{Frequency relation}

\begin{figure}[t]
    \centering \includegraphics[width=0.8\textwidth]{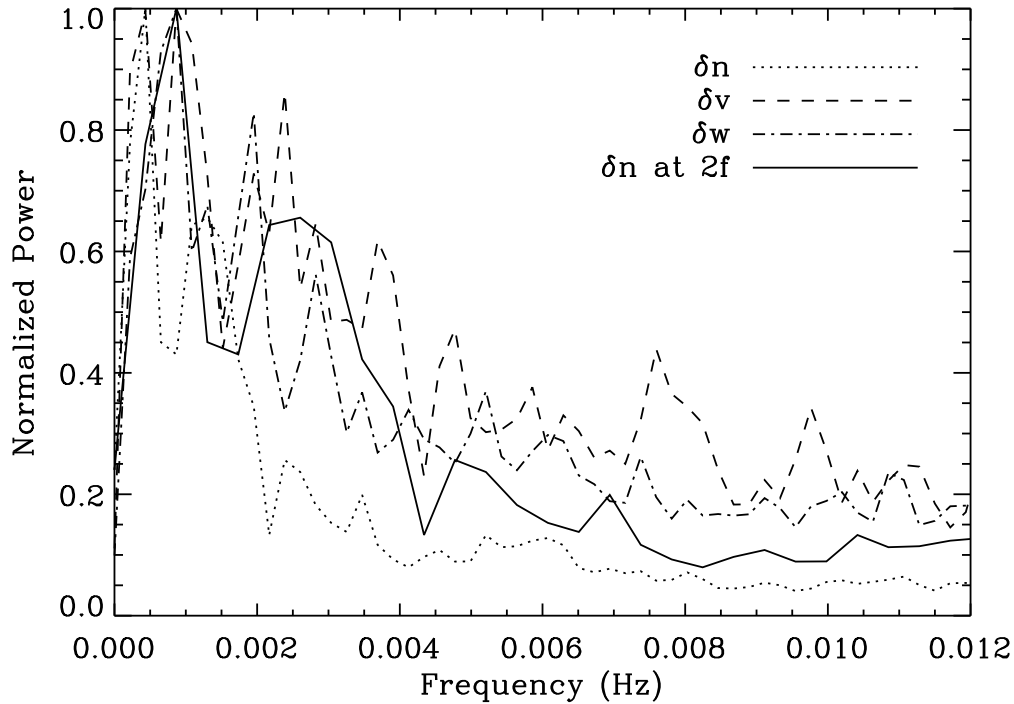}
    \caption{\label{fig:divsdv} Average power spectrum for the density fluctuations (labeled $\delta n$; dotted curve), velocity fluctuations (labeled $\delta v$; dashed curve), and line width fluctuations (labeled $\delta w$; dash-dotted curve), normalized to their respective maximum values. The solid curve shows the average power spectrum for the density fluctuations when its frequency axis is multiplied by a factor of two. In that case, the peaks $\delta n$ power spectrum align better with peaks in the $\delta v$ and $\delta w$ power spectra. The relative uncertainties in the normalized power are estimated to be about 13\%. 
    }
\end{figure}

The power spectrum for the density fluctuations, $P_{\mathrm{\delta n}}(f)$, resembles the power spectrum for the velocity fluctuations, $P_{\mathrm{\delta v}}(f)$, and they appear to differ only by a scaling of the frequency axis (Figure~\ref{fig:divsdv}). Note that since these power spectra were derived from the same number of underlying data points and were analyzed in the same way, the frequency axes are identical. In order to quantify the frequency scaling, we calculated a correlation coefficient between the two power spectra as a function of a scaling factor applied to the $P_{\mathrm{\delta n}}(f)$ frequency axis. To calculate this correlation, we multiply the frequency axis for $P_{\mathrm{\delta n}}(f)$ by a scaling factor, $\alpha$ to obtain $P_{\mathrm{\delta n}}(\alpha f)$, which expands or contracts the power spectrum. Since the frequency scale $f$ is discrete, we would like the $\alpha f$ frequency values to match up with $f$ frequency values, so it is necessary to interpolate $P_{\mathrm{\delta n}}(\alpha f)$ back to the original $f$ axis. Once that is done, we compute the correlation coefficient, $c$, between the scaled $P_{\mathrm{\delta n}}$ and original $P_{\mathrm{\delta v}}$ spectra using the standard formula \citep[e.g.,][]{Jenkins:1968}
\begin{equation}
c=\frac{\sum_{i=0}^{N} (x_i-\bar{x})(y_i-\bar{y})}
{\sqrt{\sum_{i=0}^{N} (x_i-\bar{x})^2 \sum_{i=0}^{N}(y_i-\bar{y})^2}}, 
\label{eq:defcor}
\end{equation}
where $x$ and $y$ represent the two quantities being compared, $\bar{x}$ and $\bar{y}$ are their average values, and $N$ is the total number of data points. This correlation coefficient was repeated for a range of $\alpha$ values. 

\begin{figure}[t]
	\centering \includegraphics[width=0.8\textwidth]{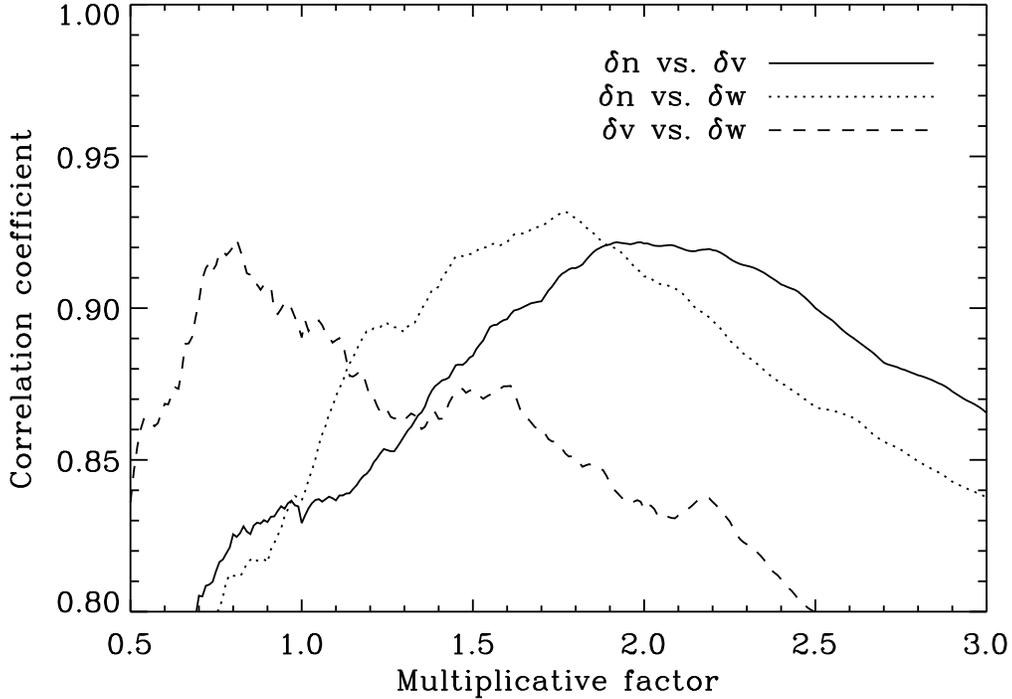}
	\caption{\label{fig:scaling} Correlation coefficient versus the scaling factor for the frequency axis for the power spectra of $\delta n$ vs. $\delta v$ (solid curve), $\delta n$ vs. $\delta w$ (dotted curve), and $\delta v$ vs. $\delta w$ (dashed curve). In each case, it is the first variable listed whose frequency axis is to be multiplied. 
	}
\end{figure}

The solid curve in Figure~\ref{fig:scaling} shows the correlation coefficient between the power spectra for the $\delta n$ and $\delta v$ fluctuations as a function of the density power spectrum frequency scaling factor. Figure~\ref{fig:scaling} shows the results for the Si~\textsc{iv} 1394~\AA\ line, but we also performed the same analysis for the 1403~\AA\ line. In order to check the sensitivity to our averaging procedure, we also performed the same analysis on power spectra where we weighted the spatial averaging by the statistical significance of the peaks. In all cases, we found similar results. Based on the average peak location in all of the $c(\alpha)$ plots, we found $\alpha = 2.01 \pm 0.12$. 


\subsection{Line Widths and Line Width Fluctuations}\label{subsec:width}

The width, $w$, of a spectral line is set by instrumental broadening, thermal broadening, and non-thermal broadening $w = \sqrt{ w_{\mathrm{inst}}^2 + w_{\mathrm{th}}^2 + w_{\mathrm{nt}}^2}$. Throughout, all widths refer to the $1\sigma$ Gaussian width. The median Si~\textsc{iv} line width measured by \textit{IRIS} in the off-limb region studied here was $w = 0.12$~\AA. \textit{IRIS} has $w_{\mathrm{inst}}=0.011$~\AA\  \citep{DePontieu:SolPhys:2014}. The peak formation electron temperature for Si~\textsc{iv} is $T_{\mathrm{e}} = 8\times10^{4}$~K \citep{Dere:ApJS:2019}. Assuming that the ion temperature is equal to the electron temperature, the corresponding estimated Si~\textsc{iv} thermal width is $w_{\mathrm{th}}=0.023$~\AA. 
Thus, the \textit{IRIS} line width is dominated by $w_{\mathrm{nt}}$. This non-thermal broadening is caused by unresolved fluid motions along the line-of-sight, such as due to waves. After subtracting the estimated instrumental and thermal width, and converting to velocity units, we find that the median $w_{\mathrm{nt}} = 24.2$~$\mathrm{km\,s^{-1}}$. This can be interpreted as an estimate of the Alfv\'enic wave amplitude. Using a line width to estimate the wave amplitude is more reliable estimate than using the Doppler shift, which only provides a lower limit for the wave amplitudes, since the broadening is not washed out by the line of sight integration \citep{McIntosh:ApJ:2012}. On the other hand, other flows can also contribute to $w_{\mathrm{nt}}$, so one can conservatively consider $w_{\mathrm{nt}}$ to be an upper bound for the wave amplitudes. 

\begin{figure}[t]
	\centering \includegraphics[width=0.8\textwidth]{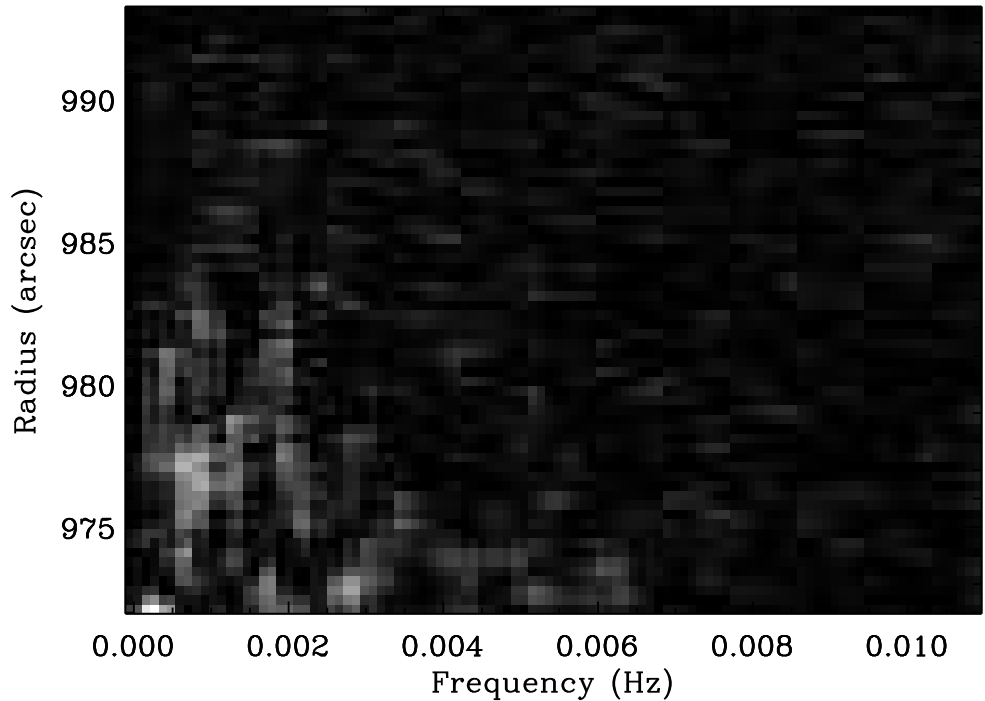}
	\caption{\label{fig:Wpsd_vs_y} Same as Figure~\ref{fig:Ipsd_vs_y}, but for the line width fluctuations. The color scale is linear with a maximum value of 0.13. 
	}
\end{figure}

We also observed fluctuations in the line width around the average value discussed above. The power spectrum of these $\delta w$ fluctuations was analyzed using the same methods as for intensity and wavelength. Figure~\ref{fig:Wpsd_vs_y} shows the $\delta w$ power spectrum as a function of height, which has a frequency distribution that resembles that of the $\delta \lambda$ fluctuations shown in Figure~\ref{fig:Vpsd_vs_y}. Figure~\ref{fig:scaling} illustrates the correlation analysis for the frequency scaling, which shows that the frequency scaling between the $\delta v$ and $\delta w$ fluctuations is 0.81, while the frequency scaling for $\delta w$ versus $\delta n$ fluctuations is 1.8. So, the $\delta v$ and $\delta w$ fluctuations have a similar power spectra, while the power spectra for the $\delta n$ and $\delta w$ fluctuations appear to have the same nearly factor of two frequency scaling that we previously found for the $\delta n$ versus $\delta v$ fluctuations. The RMS amplitude of the $\delta w$ fluctuations was about $0.01$~\AA, which corresponds to a velocity of $2.2$~$\mathrm{km\,s^{-1}}$. This is very similar to the $0.012$~\AA\ amplitude of the Doppler shift fluctuations. 

All of these results suggest that $\delta w$ and $\delta \lambda$ measure the same underlying fluctuations. One possible explanation is that they are both independently showing properties of the Alfv\'enic waves in the transition region. The $\delta \lambda$ Doppler shift fluctuations might show oscillations back and forth along the line of sight, while the $\delta w$ fluctuations may show broadening from structures where the back and forth motions are not resolved within a single pixel, for example due to unresolved small rotating structures. Another possibility is that the relation between $\delta w$ and $\delta \lambda$ is due to multiple flow components along the line of sight that are not resolved in our single-Gaussian fits. For example, if much of the plasma is stationary, but a fraction is Doppler shifted, the actual line would be a large Gaussian peak blended with a smaller slightly shifted Gaussian. When a single-Gaussian model is fit to such a line profile, the result will be an apparent Doppler shift in the line centroid and a slightly broader line width. Inspection of the fits did not reveal any clear evidence for two-Gaussian functions, but as the Doppler shifts are small compared to the line widths any secondary components may be difficult to resolve. 
From a pragmatic perspective, the $\delta w$ fluctuations do not provide any information that is not already less ambiguously provided by the $\delta \lambda$ analysis. So, it is not necessary to resolve the issue here. 

\subsection{Fluctuation Speeds}

\subsubsection{Density Fluctuations}

We measured the propagation speed of the density fluctuations using both time-domain cross-correlation and frequency-domain cross-spectrum techniques. For the cross-correlation approach, we computed the cross-correlation function for $\delta I(t,y_i)$ at each height $y_i$, with all larger heights $\delta I(t,y_j)$. The lag-time $\tau_{ij}$ at which the cross correlation peaks represents the travel time of the wave to go from $y_i$ to $y_j$. We quantify $\tau_{ij}$ as the first moment of the peak in the cross correlation and took the uncertainty to be the second moment of the peak. This gives us the lag time as a function of the height difference, $\tau_{ij}(\Delta y_{ij})$. For each initial position $i$, we perform a linear least squares fit to $\tau_{ij}$ versus $\Delta y_{ij}$, whose slope represents the propagation speed. We found that the cross correlation loses coherence after about 5~Mm, so we performed the fit over a height range of 4.5~Mm. From this method we found the density fluctuation velocity was $v_{\mathrm{\delta n}} = 75 \pm 10$~$\mathrm{km\,s^{-1}}$. 

We also used the cross-spectrum method to estimate the velocity (i.e., phase speed) of the density fluctuations \citep{Athay:ApJS:1979a}. The cross spectrum gives the phase difference $\Delta \phi_{ij}(f)$ of the fluctuations between two heights $y_i$ and $y_j$ as a function of the frequency $f$. The time delay is then $\Delta \phi_{ij}(f)/(2 \pi f)$. By computing the cross spectrum between each pair of heights, we inferred the delay time between those heights and then perform a linear fit to find the velocity, just as we did for the cross correlation. The advantage of the cross-spectrum method is that it can, in principle, measure dispersion in the fluctuations, i.e., identify whether the phase speed depends on the frequency. A disadvantage is that resolving the data along the frequency dimension increases the statistical uncertainty. For these data, the uncertainties were large enough that we cannot conclude anything about dispersion. Instead, we focus on a couple of peaks in the power spectrum (Figure~\ref{fig:Ipsd_vs_y}) to find the velocity.  We found that for the peak at $f=5.9$~mHz $v_{\mathrm{\delta n}} = 75.4 \pm 6.3$~$\mathrm{km\,s^{-1}}$ and for $f=1.08$~mHz $v_{\mathrm{\delta n}}=97 \pm 43$~$\mathrm{km\,s^{-1}}$. 

Based on these two methods, the phase speed for the density fluctuations was found to be about $75$~$\mathrm{km\,s^{-1}}$. This can be compared to the theoretically expected sound speed,  $c_{\mathrm{s}} = \sqrt{2\gamma k_{\mathrm{B}}T_{\mathrm{e}}/m_{\mathrm{i}}}$, where $\gamma=5/3$ is the adiabatic index, $k_{\mathrm{B}}$ is the Boltzmann constant, and $m_{\mathrm{i}}$ is the average ion mass. We take $m_{\mathrm{i}}=1.15$~$m_{\mathrm{p}}$, with $m_{\mathrm{p}}$ the proton mass, in order to account for a 5\% helium concentration. This expression for $c_{\mathrm{s}}$ also assumes that the ion temperature is equal to the electron temperature, which we estimate as the Si~\textsc{iv} formation temperature of $8\times10^{4}$~K, the expected sound speed is $c_{\mathrm{s}}=44$~$\mathrm{km\,s^{-1}}$. So our measured $v_{\mathrm{\delta n}}$ is similar too, though larger than, the expected sound speed. One possible explanation for the discrepancy is that the transition region may be multithermal or not in ionization equilibrium, so that the actual temperature could be larger than our rough estimate based on the Si~\textsc{iv} formation temperature. 


\subsubsection{Velocity Fluctuations}

\begin{figure}[t]
	\centering \includegraphics[width=0.8\textwidth]{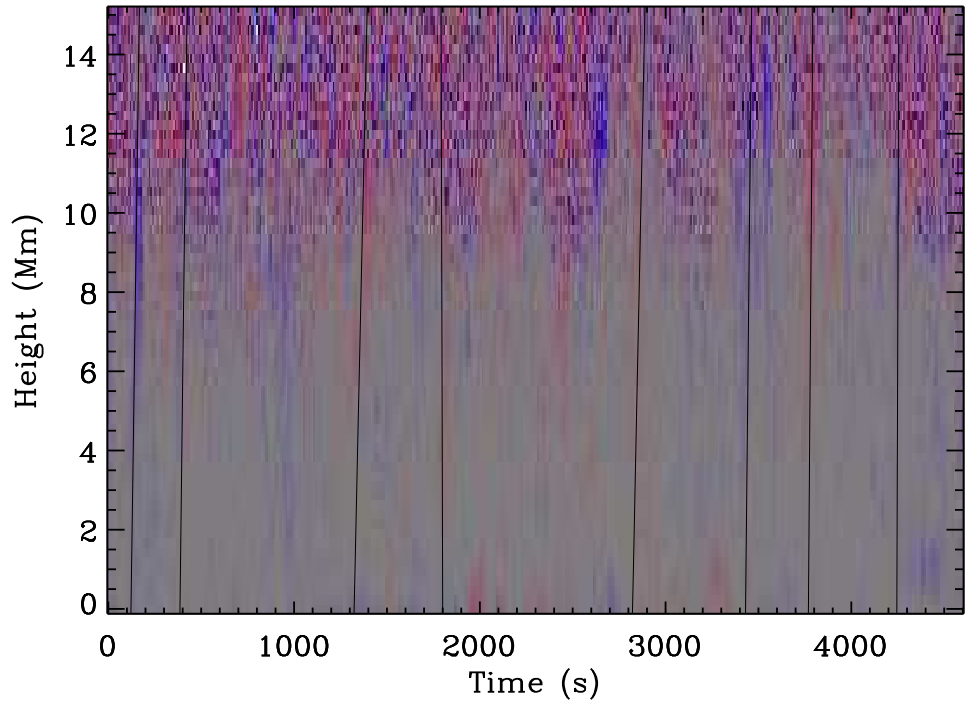}
	\caption{\label{fig:dvspeed} Time-distance plot of $\delta \lambda$ fluctuations, which are proportional to $\delta v$ fluctuations. The color scale is linear from $\pm 0.34$~\AA, with red negative, blue positive, and grey zero. Black lines show the fit to tracked moving structures in the time-distance plot. The slopes of these lines indicate the speed. 
	}
\end{figure}

The velocity fluctuations propagate much faster than the density fluctuations, so that they appear as nearly vertical lines in a height-time diagram. The very small travel times or phase delays between two heights and the uncertainties in the data, precluded a cross-correlation or cross-spectrum analysis. Instead, we quantified the phase speed by tracking the motion of features in the height-time plot (Figure~\ref{fig:dvspeed}). Moving features were initially identified by eye. Then, the time at which the feature was present at each height was determined by finding the centroid position (first moment) of the feature along the time axis. We performed a linear fit to the centroid time versus height in order to find the velocity of each identified feature. Based on 8 identified features, we found a weighted mean speed of the velocity fluctuations of $v_{\mathrm{\delta v}} = 250 \pm 20$~$\mathrm{km\,s^{-1}}$. One should be cautious about taking this number to be too precise, due to the potential for systematic bias in the feature identification step. 

This estimated speed is similar in magnitude to what is expected for Alfv\'en waves. The Alfv\'en speed is $V_{\mathrm{A}} = B/\sqrt{4 \pi \rho}$, where $B$ is the magnetic field strength and $\rho$ is the mass density. We take the mean particle mass (including electrons and ions) to be $0.6 m_{p}$ in the corona. Then for $B \approx 4$~G and $n_{\mathrm{e}}=2.6 \times 10^{9}$~$\mathrm{cm^{-3}}$ (as described in the next section), we estimate $V_{\mathrm{A}} \sim 220$~$\mathrm{km\,s^{-1}\,s^{-1}}$. 


This estimate also confirms the suspected problem with the short travel times through the limited observed height range. As the vertical pixels span a height of 0.24 Mm per pixel and the cadence is $\Delta t$ =9.33~s, the theoretical Alfv\'en speed implies that the Alfv\'en waves move at a speed of about 7.3 vertical pixels per frame. The measured height range spans 64 vertical pixels, but the useful height range is more strongly limited by noise level. As a result, Alfv\'en waves are nearly vertical lines in the height-time plot, as we observe. We can say with confidence that the velocity fluctuations are significantly faster than the density fluctuations. 

The empirical value for the Alfv\'en speed allows us to estimate the amplitude of the Alfv\'en wave relative to the average magnetic field, $\delta B/B_{0}$. For an Alfv\'en wave $\delta v/V_{\mathrm{A}} = \delta B/B_{0}$. Thus, the estimated range for $\delta B/B = 0.01$--$0.1$, where the minimum value is based on the Doppler shift fluctuations and the maximum value is based on the average line width.

\subsection{Plasma properties}

We measured the density using the ratio of two O~\textsc{iv} lines \citep{Polito:AA:2016}. Specifically we used the ratio of the $2s^2\,2p\, ^{2}P_{1/2} - 2s\,2p^2\, ^{4}P_{1/2}$ transition at 1399.78~\AA\ to $2s^2\,2p\, ^{2}P_{3/2} - 2s\,2p^2\,^{4}P_{5/2}$ transition at 1401.16~\AA. There is a weak photospheric line in the wing of the 1401.16~\AA\ line, but in the off-limb data both lines are unblended. 

In order to use the O~\textsc{iv} line intensity ratio as a diagnostic we first performed an absolute calibration of the \textit{IRIS} data to convert from DN to intensity units \citep{IRIS:StellarCalibration, Wulser:SolPhys:2018}. The intensities were then extracted from the spectrum in two ways. In one method, we fit Gaussian line profiles to the O~\textsc{iv} lines in order to extract the amplitude, line width, and centroid. However, for low intensities the data is noisy and the fits may be less reliable. To address that problem, we also integrated the spectrum directly to obtain the total intensity. Both methods resulted in similar intensity ratios and densities. To interpret the intensity ratio as a density, we used the CHIANTI atomic database \citep{Dere:AAS:1997, Dere:ApJS:2019}. 

Density measurements were most consistent in the height range between 970--980$^{\prime\prime}$. Presumably because the lines were brightest at those lowest heights and the density there lies within the range at which the diagnostic is most sensitive. We found that the weighted mean density was $\log_{10}(n_{\mathrm{e}}) = 9.42 \pm 0.05$ based on the Gaussian fit method and $\log_{10}(n_{\mathrm{e}}) = 9.42 \pm 0.02$, based on the direct integration method. That is, $n_{\mathrm{e}} = 2.6 \pm 0.3 \times 10^{9}$~$\mathrm{cm^{-3}}$. 

A potential systematic uncertainty is that the O~\textsc{iv} formation temperature of $1.4\times10^{5}$~K is somewhat higher than that of Si~\textsc{iv} at $8\times10^{4}$~K. There is not a sufficient range of ion species to perform a detailed temperature analysis, and the transition region is probably multi-thermal. Our results do not depend sensitively on the temperature, so wherever an estimate is needed we take $T_{\mathrm{e}} \approx 8 \times 10^{4}$~K. 

For comparing to theory, it is useful to estimate the plasma $\beta$, which is given by $\beta = 8\pi n_{\mathrm{e}} k_{\mathrm{B}} T_{\mathrm{e}}/B^2$. Using our inferred density and estimated temperature, we find $\beta = 0.73/B^2$, where $B$ is the magnetic field strength in Gauss. 
We do not have a direct measurement of the polar magnetic fields. However, published estimates suggest that the magnetic field strength is in the range of 2--6~G \citep{Tian:AA:2008, Wang:ApJ:2009, Janardhan:AA:2018}. The corresponding range is $\beta \approx 0.02$--$0.18$. 

\section{Comparison of Observations with PDI Theory}\label{sec:discuss}

The apparent frequency-scaling relationship between the density and velocity fluctuations suggests that these oscillations are coupled to one another. One possibility is that this interaction occurs via PDI. Here, we discuss the observations in this context and show that they are consistent with theoretical predictions for PDI. 

In PDI, a forward propagating pump Alfv\'en wave decays into a forward sound wave and a backward Alfv\'en wave. Our observations are consistent with seeing the pump wave and the secondary sound wave. Based on their speeds, the observed fluctuations are consistent with the density fluctuations representing sound waves and the velocity fluctuations being Alfv\'en waves. We found that the density fluctuations propagated at about 75~$\mathrm{km\,s^{-1}}$, which is similar to the expected sound speed. The velocity fluctuations propagated at about 250~$\mathrm{km\,s^{-1}}$, which is about the expected Alfv\'en speed. The expected backward propagating secondary Alfv\'en wave is not clearly seen, but it may be obscured because of its smaller amplitude and because the direction of propagation is difficult to resolve given the fast phase speed and limited height range. 

\begin{figure}[h]
	\centering \includegraphics[width=0.9\textwidth]{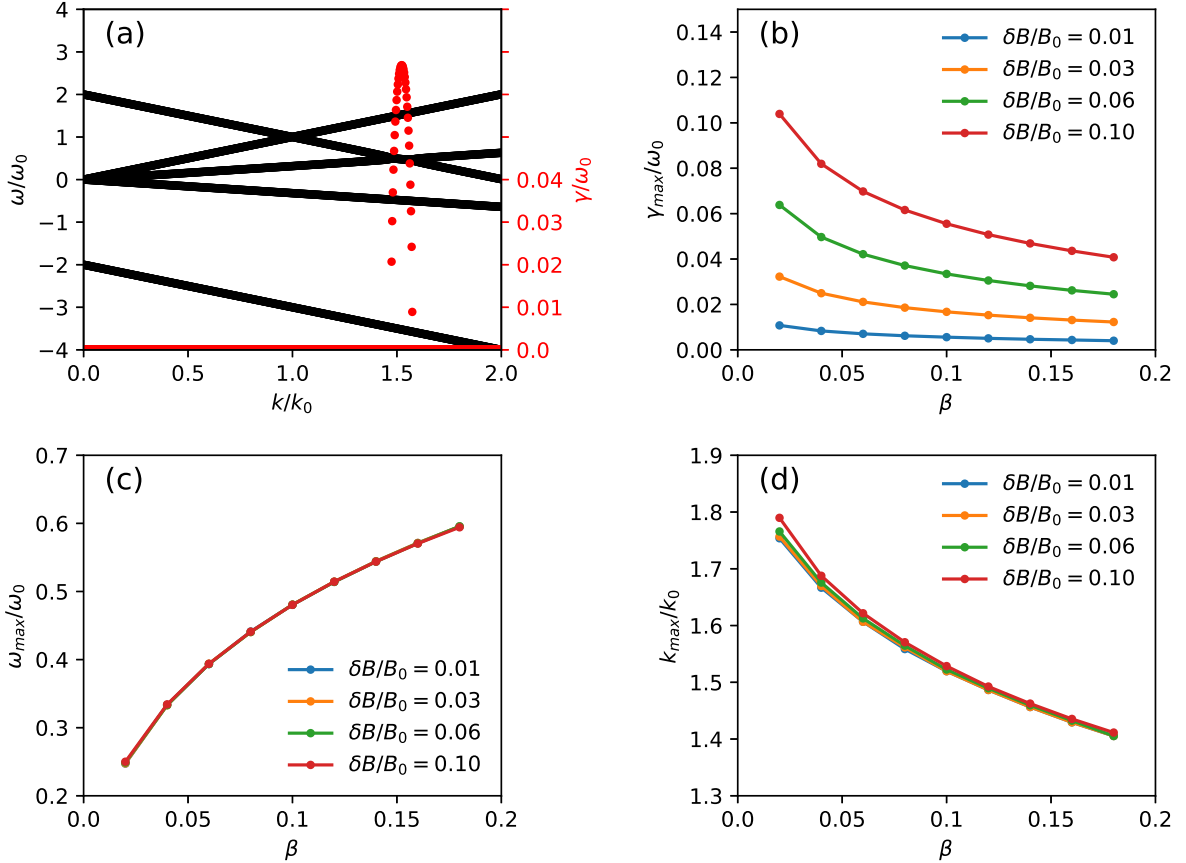}
	\caption{\label{fig:pdi} 
	(a) Numerical solutions to the nonlinear dispersion relation for parametric decay instabilities of an Alfven wave \citep{derby_apj_1978} with $\delta B/B_0=0.06$ and $\beta=0.1$. The black dots, which blend together to appear as black lines, show real frequencies $\omega$, and the red dots show growth rates $\gamma$, both of which are normalized to the frequency of the pump Alfven wave $\omega_0$. The wavenumbers are normalized to the wavenumber of the pump wave $k_0$ (note that $\omega_0=k_0 V_A$). At any given $k$ there are five solutions, most of which are stable (i.e. $\gamma=0$). The unstable solution near $k/k_0=1.53$ corresponds to PDI, with a maximum growth rate $\sim 0.032$, and frequency $\sim 0.48$. (b) The growth rate, (c) the frequency, and (d) the wavenumber of the most unstable (largest $\gamma$) PDI mode for various combinations of $\delta B/B_0$ and $\beta$. All three quantities depend on $\beta$, but $\omega_{max}$ and $k_{max}$ have little dependence on $\delta B/B_0$ (In (c) four curves lie on top of one another). 
	}
\end{figure}	

In order to make a more detailed comparison with the predictions of the theory of PDI, we have numerically solved the dispersion relation for PDI from the theory of \citet{derby_apj_1978}. This theory is based on magnetohydrodynamics (MHD) with a uniform background magnetic field, $B_0$, and $\beta$ not necessarily small. The equilibrium state of the plasma also includes a circularly polarized pump Alfv\'en wave of finite amplitude $\delta B$, frequency $\omega_0$, and wavenumber $k_0$, related by $V_{\mathrm{A}}=\omega_0/k_0$. The dispersion relation for PDI is found by linearizing the MHD equations for parallel propagating fluctuations from this equilibrium state. This dispersion relation is given in Equation~(17) of \citet{derby_apj_1978}, which is a  nonlinear algebraic equation involving a fifth order polynomial of $\omega$ and $k$, which are the frequency and wavenumber of the daughter mode, respectively. There are two parameters, $\beta$ and $\delta B/B_0$, in the dispersion relation.

For discrete values of real $k$, we solve the dispersion relation numerically to obtain solutions for complex $\omega$. The real part of $\omega$ is the real frequency of the daughter mode, while the imaginary part, $\gamma$ is the exponential growth rate for the amplitude of the mode. The growth rate for PDI depends on the two parameters, $\delta B/B_0$ and and $\beta$ \citep{derby_apj_1978, golds_apj_1978}. Figure~\ref{fig:pdi}(a) shows one example of this solution for $\delta B/B_0 = 0.06$ and $\beta = 0.1$, which are plausible parameters based on the observations. Figures~\ref{fig:pdi}(b-c) are found by solving the dispersion relation for a range of other values of $\delta B/B_0$ and $\beta$. 

 Figure~\ref{fig:pdi}(b) shows the maximum growth rate for PDI for various combinations of $\delta B/B$ and $\beta$ and demonstrates that the growth rate is largest for large pump wave amplitudes $\delta B/B_0$ and low $\beta$. This is also consistent with the theory of \citet{Sagdeev:1969}, which for the limit of small amplitudes and small $\beta$ predicts $\gamma/\omega_0 \approx (1/2) (\delta B/B_0) \beta^{-1/4}$. 

Figures~\ref{fig:pdi}(c) and (d) plot the frequency $\omega_{\mathrm{max}}$ and wavenumber $k_{\mathrm{max}}$, for which the maximum instability growth rate occurs. In both cases, these quantities are normalized by the frequency and wavenumber for the pump wave. Panels (c) and (d) show that $\omega_{\mathrm{max}}$ and $k_{\mathrm{max}}$ depend almost entirely on $\beta$ and not on the wave amplitude. For low $\beta$,  $\omega_{\mathrm{max}}/\omega_0 \approx 1/2\sqrt{\beta}$ \citep{Sagdeev:1969}.

Since the sound wave is produced by PDI, it is expected that the sound waves will be excited most strongly at the frequency $\omega_{\mathrm{max}}$. Observationally, we found that the ratio $\alpha = \omega_{0}/\omega_{\mathrm{max}} = 2.01 \pm 0.12$. Figure~\ref{fig:pdi}(c) indicates that this corresponds to $\beta \approx 0.1$. Our measured and estimated values for density, temperature, and magnetic field imply a range for $\beta = 0.02$--$0.18$. So, the $\beta$ implied by the theory of the PDI based on the ratios of the observed fluctuation frequencies is in the middle of the estimated range based on our independent estimates of plasma properties. 



Assuming the fluctuations are driven by PDI, we can estimate the growth rate from the measured wave amplitudes. The Alfv\'en wave amplitude can be expressed in terms of the velocity as $\delta B/B_0 = \delta v/V_{\mathrm{A}}$. The RMS amplitude of the velocity fluctuations from Doppler shift was 2.6~$\mathrm{km\,s^{-1}}$, which combined with $V_{\mathrm{A}} \approx 250$~$\mathrm{km\,s^{-1}}$ gives an Alfv\'en wave amplitude of $\delta B/B_0 \approx 0.01$. This is a lower bound since the line-of-sight integration tends to wash out the Doppler shifts. Using the average line width instead, we inferred a non-thermal velocity of $v_{\mathrm{nt}} = 24.2$~$\mathrm{km\,s^{-1}}$, corresponding to $\delta B/B_0 \approx 0.1$. This is an upper bound, because  the non-thermal velocity can be influenced by flows other than waves. As shown in Figure \ref{fig:pdi}(b), the maximum growth rate of PDI is in the range from $0.005\omega_0$ to $0.10\omega_0$, depending of the wave amplitude and plasma beta. Taking observationally plausible values of $\beta=0.1$ and $\delta B/B_0=0.06$, the growth rate of the PDI is about $0.035\omega_0$. In absolute units, this is $2\times 10^{-4}$~$\mathrm{s^{-1}}$ for a pump Alfven wave  with a frequency of $10^{-3}$~$\mathrm{s}^{-1}$.

Although the observational evidence is consistent with PDI, the slow growth rate of the instability and the possible presence of large gradients in this region raise a challenge for this interpretation. For an $V_{\mathrm{A}} = 250$~$\mathrm{km\,s^{-1}}$ and growth rate $\gamma = 2\times 10^{-4}$~$\mathrm{s^{-1}}$, the wave propagation length during a growth time is about $10^{3}$~Mm, of order $\sim R_{\sun}$. The temperature, density, and magnetic field gradients in the corona are expected to have length scales shorter than this. Under these conditions, we expect the properties of PDI to be modified from the linear theory used above. 

There are few theoretical studies of PDI in an inhomogeneous plasmas. \citet{Tenerani:JGR:2013} have studied PDI numerically, using an expanding box model to discern the effects of solar wind expansion on the PDI. They found that expansion tends to reduce the growth rate, because the resonance condition changes as the solar wind flows outward. Numerical studies under conditions relevant to the lower solar atmosphere are needed in order to understand the effects of inhomogeneity in the observed region. 

We should consider whether there are alternative interpretations of the observations other than PDI. One possibility is that the relation between the density and velocity fluctuations is the manifestation of some underlying wave mode having a velocity fluctuation at a harmonic of the compressional oscillation. There is strong evidence against this possibility since such a mode should have a single phase speed so that both the velocity and density fluctuations should move at the same speed, whereas here we find that the velocity fluctuation is significantly faster than the density fluctuation. Still, the velocity fluctuation speed is not measured as precisely or systematically as the other quantities discussed here, which is a systematic uncertainty that future measurements should aim to resolve. 

It is also possible that the Alfv\'enic waves and density fluctuations are not directly coupled, but that the frequency relation arises as part of an unknown process that would excite both types of waves at lower heights. This possibility could potentially be distinguished from PDI by observing the direction of propagation of the Alfv\'en waves. PDI predicts both upward and downward propagating Alfv\'en waves, but in the absence of PDI we expect the waves to be propagating upward only. This assumes that there are no other sources of reflection and downward propagating waves besides PDI. Our analysis was not able to resolve this issue. The structures identified for the velocity fluctuation speed analysis were predominantly upward, presumably representing the pump Alfv\'en waves, although the height-time diagram also appears to show some downward moving fluctuations. In fact, the dominance of upward Alfven waves in the process of PDI was also observed in numerical simulations \citep{Li:ApJ:2022}. To measure the power in both upward and downward propagating waves systematically, one could apply a two dimensional Fourier analysis to the height-time diagram to generate a wavenumber-frequency ($k-\omega$) diagram \citep[see e.g.,][]{Tomczyk:ApJ:2009}. Such an analysis was not possible for this data set due to the noise level and the limited height range, which limits the minimum resolvable wavenumber. 


\section{Conclusions}\label{sec:conclusions}

We find evidence for the first observation of PDI in the lower solar atmosphere near the transition region. Our analysis shows that the power spectrum of density fluctuations matches the power spectrum of the velocity fluctuations except for a scaling of the frequency axis. This scaling factor matches the frequency relationship between a pump Alfv\'en wave and the secondary sound wave it is theoretically expected to drive via PDI given the estimated plasma $\beta$ estimated for the observed region. 

Other possible processes that could explain the observed relation between the density and velocity fluctuations are that they are due to the same underlying wave mode or that the relation arises due to nature of the wave excitation process at lower heights. The wave mode hypothesis is unlikely, as the density and velocity fluctuations appear to be propagating at very different speeds. An explanation in terms of some unknown wave excitation process cannot be ruled out based on these observations. That possibility could be resolved by better quantifying the power in upward versus downward waves in future work. Developments in theory are also needed to understand how the inhomogeneity of the plasma properties in this region affect the properties of the PDI. 

The observation of PDI in the transition region of the Sun has implications for broader models of turbulence and coronal heating. The observed region studied here is a fairly generic observation of the transition region at the base of a coronal hole, so it is likely that PDI is ubiquitous in such regions. This would support numerical models for coronal heating and solar wind acceleration \citep[e.g.,][]{Shoda:ApJ:2016} that suggest PDI as a mechanism for promoting turbulence and heating. Future work should look for PDI in other structures. If PDI is indeed common, then PDI may be a fundamental process in the Sun that mediates the transfer of energy into the corona. 


\begin{acknowledgments}

M. H. and D. W. S acknowledge support from NSF STR Grant 1384822 and NASA LWS Grant 80NSSC20K0183. X. F. is supported by NASA LWS Grant 80NSSC20K0377. 

\end{acknowledgments}

\clearpage

\bibliography{PDI}

\end{document}